\documentclass[twocolumn,showpacs,preprintnumbers,floats,amsmath,amssymb]{revtex4}
\usepackage{graphicx}
\usepackage{dcolumn}
\usepackage{bm}
\begin{document}
\newcommand{\ds}{\displaystyle}
\newcommand{\be}{\begin{equation}}
\newcommand{\en}{\end{equation}}
\newcommand{\bea}{\begin{eqnarray}}
\newcommand{\ena}{\end{eqnarray}}
\topmargin -2cm
\title{Relic gravitons and viscous cosmologies}
\author{Mauricio Cataldo}
\altaffiliation{mcataldo@ubiobio.cl} \affiliation{Departamento de
F\'\i sica, Facultad de Ciencias, Universidad del B\'\i o--B\'\i
o, Avenida Collao 1202, Casilla 5-C, Concepci\'on, Chile.\\}
\author{Patricio  Mella}
\altaffiliation{patriciomella@udec.cl} \affiliation{Departamento de F\'{\i}sica, Universidad de Concepci\'{o}n,\\
Casilla 160-C, Concepci\'{o}n, Chile.}
\date{\today}
\begin{abstract}
{\bf {Abstract:}} Previously it was shown that there exists a
class of viscous cosmological models which violate the dominant
energy condition for a limited amount of time after which they are
smoothly connected to the ordinary radiation era (which preserves
the dominant energy conditions). This violation of the dominant
energy condition at an early cosmological epoch may influence the
slopes of energy spectra of relic gravitons that might be of
experimental relevance. However, the bulk viscosity coefficient of
these cosmologies became negative during the ordinary radiation
era, and then the entropy of the sources driving the geometry
decreases with time.

We show that in the presence of viscous sources with a linear 
barotropic equation of state $p=\gamma \rho$ we get viscous cosmological 
models with positive bulk viscous stress during all their evolution, and
hence the matter entropy increases with the expansion time. In
other words, in the framework of viscous cosmologies, there exist
isotropic models compatible with the standard second law of
thermodynamics which also may influence the slopes of energy
spectra of relic gravitons.

\vspace{0.5cm} 
\end{abstract}
\smallskip\
\maketitle

Our universe can be viewed as containing a sea of stochastically
distributed gravitational waves of primordial origin. Among
all observational cosmological evidences (present and future),
primordial gravitational waves should have a sufficiently enlightened character in
order to better understand the very early universe. The gravitational waves of
cosmological origin are nothing but squeezed states of many
gravitons produced from the vacuum fluctuations of the background
metric. A qualitative analysis can be performed in the context of
different physical frameworks, since all models for the very early
universe predict the formation of stochastic gravitational wave backgrounds. As
examples we can mention inflationary quintessential
models~\cite{Giovannini0}, inflationary models in Brans-Dicke
theory of gravity~\cite{Sahoo}, cosmological models in the
Brane-world scenario~\cite{Cavaglia}, and superstring
theories~\cite{Giovannini000}. The shape of the stochastic
graviton background spectrum is affected by the variations of the
background dynamics. 

In this context, Giovannini~\cite{Giovannini}
has considered the interesting possibility of constructing flat
Friedmann--Robertson--Walker (FRW) cosmologies endowed with a bulk
viscous stress which induces a violation of the dominant energy
condition (DEC) for a limited amount of time at an early
cosmological epoch. This kind of cosmological models may be
connected to some of the recent remarks of
Grishchuk~\cite{Grishchuk} concerning the detectability of
stochastic gravitational wave background by forthcoming interferometric detectors,
such as LIGO, VIRGO, GEO600, LISA~\cite{Ligo}. Effectively, bulk
viscous dissipative processes may influence the slopes of the
energy spectra of relic gravitons (generated at the time of
violation of the DEC) producing an increasing with frequency in a
calculable way. These slopes are crucially related to the sign of
the $\rho+p$, where $\rho$ and $p$ are, respectively, the energy
density and the pressure density of the cosmic fluid. The
requirement that one wants expanding and inflationary universes
implies that the energy density of the created gravitons cannot
increase with frequency if $\rho+p \geq 0$, i.e. if the DEC is not
violated. Unfortunately, previous models which exploit this idea
have a phase in their evolution where the matter entropy
decreases. Specifically was considered a class of solutions which
correspond to a viscous fluid with an equation of state given by
$p=-\rho$. In this model the early phase (where the DEC is
violated) is smoothly connected to a radiation dominated
evolution. Depending upon the sign of the bulk viscosity
coefficient, the entropy of the sources driving the geometry can
very well decrease~\cite{Giovannini1,Pavon}. 

Let us discuss the class of viscous cosmologies considered in Ref.~\cite{Giovannini} more in detail. 
In a flat FRW background the Einstein field equations in the
presence of the bulk viscosity coefficient $\xi$ can be written as
\begin{eqnarray}\label{tt}
H^2=\frac{\kappa}{3} \, \rho,
\end{eqnarray}
\begin{eqnarray}\label{rr}
H^2+ \dot{H}=-\frac{\kappa}{6} \, \left(\rho+3P_{eff} \right),
\end{eqnarray}
where the effective pressure is given by
\begin{eqnarray}\label{pprime}
P_{eff}=p-3 \xi H.
\end{eqnarray}
In this case $\kappa=8 \pi G$, $H=\dot{a}/a$, $a(t)$ is the scale
factor of the flat FRW metric and the overdot represents
derivation with respect to the cosmic time coordinate. In order to
have the notation of the paper~\cite{Giovannini} we  must identify
$M_p^2=3/\kappa$ and $p^{\prime}=P_{eff}$.

The equations~(\ref{tt})--~(\ref{pprime}) imply the energy balance
\begin{eqnarray}\label{cons}
\dot{\rho}+ 3 H (\rho+P_{eff})=0.
\end{eqnarray}
In order to have a cosmological model whose evolution violates the
DEC only for a finite amount of time, in Ref.~\cite{Giovannini} it
is assumed that
\begin{eqnarray}\label{sup}
\kappa \xi=\frac{2}{3} \frac{\dot{H}}{H}.
\end{eqnarray}
This parametrization is very reasonable since the amount of
violation of DEC is proportional to $\dot{H}$. Effectively, from
Eqs.~(\ref{tt}) and~(\ref{cons}) we have that for any solution
\begin{eqnarray}\label{DECViol}
\rho+P_{eff}=-\frac{2}{\kappa} \dot{H},
\end{eqnarray}
and then a violation of the DEC implies that $\dot{H}>0$.

\begin{figure}
\includegraphics{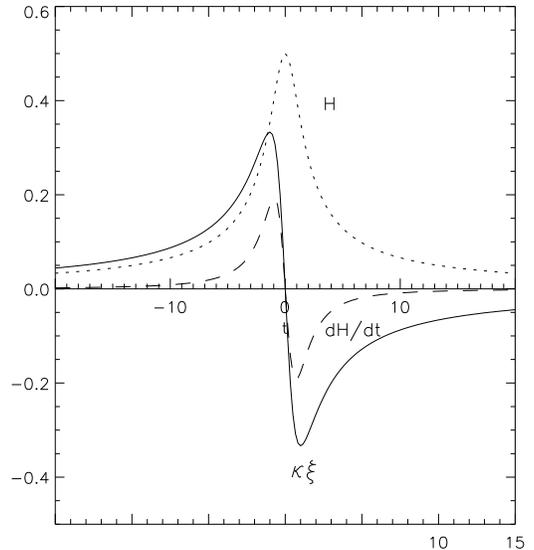}
\caption{\label{fig15} We show the behavior of $H$, $\dot{H}$ and
$\kappa \xi$ for viscous model~(\ref{GS}) and~(\ref{GS1}). We have
chosen $t_1=1$. We can see that $\dot{H}>0$ ($\xi>0$) for $t<0$
(DEC is violated) and $\dot{H}<0$ ($\xi <0$)  for $t>0$ (DEC is
preserved).}
\end{figure}

In order to have a cosmology whose early phase 
(where the DEC is violated) is smoothly connected to a radiation 
dominated evolution, Giovannini considers the scale factor given by
\begin{eqnarray}\label{GS}
a(t)= \left( t+\sqrt{t^2+t_1^2}\right)^{1/2}, 
\end{eqnarray}
and then the self-consistent solution takes the form (see Fig.\ref{fig15})
\begin{eqnarray}\label{GSas}
H=\frac{1}{2\sqrt{t^2+t_1^2}},
\end{eqnarray}

\begin{eqnarray}\label{GS1}
\kappa \xi(t)=- \frac{2 t}{3 \left(t^2+t_1^2\right)}, \,\,\,\,\,
\kappa \rho(t)=\frac{3}{4 \left( t^2+t_1^2 \right)}.
\end{eqnarray}
One can immediately see by taking the limit $t \rightarrow \pm
\infty$ that $\xi_{- \infty}(t) >0$, $\xi_{+ \infty}(t) <0$ and
$a_{\pm \infty}(t) \rightarrow (\pm t)^{\pm1/2}$. So $a(t)$ at the
final phase of the whole evolution has the behavior of the
radiation dominated era.

\begin{figure}
\includegraphics{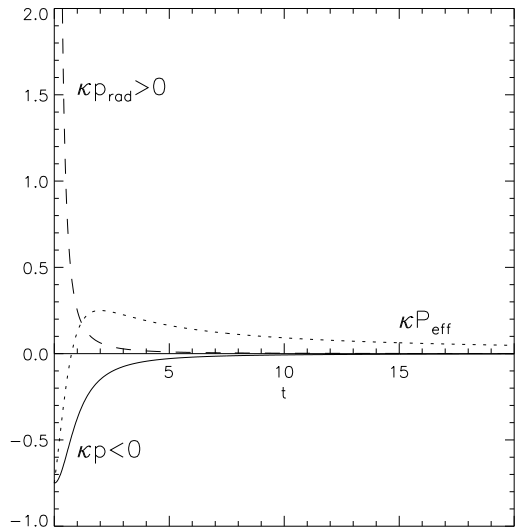}
\caption{\label{fig1} We show the behavior of the local
equilibrium pressure $p=-\rho$ (see Eq.~(\ref{GS1})), effective
pressure $P_{eff}$ (see Eq.~(\ref{PeffG})) and the radiation
pressure $\kappa p_{_{rad}}=1/4t^2$ for $t \geq 0$. We have chosen
$t_1=1$. We can see that $p$ is always negative and $P_{eff}$
behaves like $p_{_{rad}}$ for $t \rightarrow \infty$.}
\end{figure}

Now, if we put the expression~(\ref{sup}) into Eq.~(\ref{DECViol})
we conclude that the Giovannini parametrization implies that the
local equilibrium pressure is given by $p=-\rho$~\cite{Cataldo}.
So for $t \rightarrow +\infty$ the asymptotic solution to
Eqs.(\ref{GS}) and~(\ref{GS1}) is the following exact solution of
the field equations~(\ref{tt})--~(\ref{cons}):
\begin{eqnarray}
a(t)=t^{1/2}, \,\,\,\,\, \kappa \xi=-\frac{2}{3t}, \,\,\,\,\,
\kappa p=- \kappa \rho=-\frac{3}{4t^2}.
\end{eqnarray}
Notice that Eqs.~(\ref{GS}) and~(\ref{GS1}) imply that
\begin{eqnarray}\label{PeffG}
\kappa
P_{eff}=\frac{\left(4t-3\sqrt{t^2+t_1^2}\right)}{4\left(t^2+t_1^2\right)^{3/2}},
\end{eqnarray}
and then $P_{eff} \rightarrow \rho/3$ for $t \rightarrow + \infty
$, while the local equilibrium pressure behaves always as
$p=-\rho$. This is illustrated in Fig.\ref{fig1}.

Let us now consider some physical aspects of the discussed viscous
cosmological model. The thermodynamical entropy associated with
the bulk viscosity~(\ref{GS1}) decreases for $t>0$. This implies
that, for large positive cosmic time values, the second law of
thermodynamics would be violated~\cite{Pavon}. However, as was
discussed in Ref.~\cite{Giovannini1}, this statement might be not
justified since it only takes into account the matter entropy but
not the entropy of the geometry itself. This implies that, for a
FRW background, in the framework of a well defined extension of
the second law of thermodynamics, one should include both the
entropy connected with matter and the entropy connected with the
FRW background. In this context, for positive cosmic time values,
the decrease in the entropy of the sources may be compensated by
the growth of the entropy of the FRW background.
\begin{figure}
\includegraphics{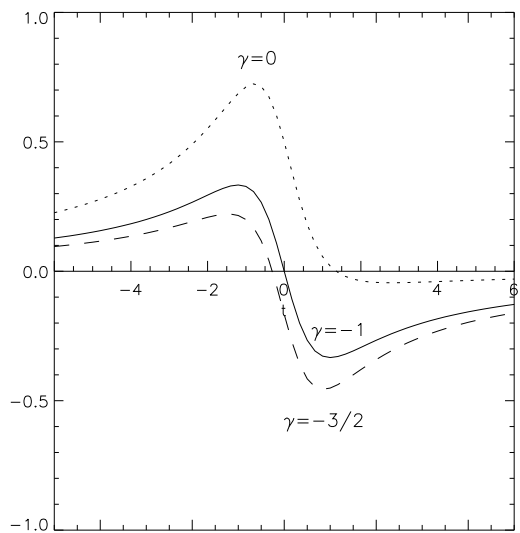}
\caption{\label{fig159} Some curves for the bulk viscous stress
given in Eq.~(\ref{CS1}) are plotted. We set $t_1=1$ and
$\gamma=-3/2,-1, 0$ as illustrative values. We can see that each
curve has a phase where it is positive, and then it transits to a
negative phase.}
\end{figure}

Unfortunately, at the present such an extension of the second law
of thermodynamics is ambiguous even for cosmological models which
do not violate the DEC (see for example~\cite{Davies}). So in this
paper we are interested in finding viscous cosmological models
which do not need such an extension of the second law of
thermodynamics.

In the following we shall generalize the Giovannini class of solutions
discussed above in order to include cosmological scenarios which always have a
positive viscous coefficient and thus satisfy the standard second
law of thermodynamics.

We will search for wider classes of solutions in the presence of viscous sources with a
linear barotropic equation of state $p=\gamma \rho$ for the local pressure. Specifically, we are 
interested in studying solutions which preserve
the form of the scale factor~(\ref{GS}) (in order to keep all
physical properties explaining the slopes of the energy spectra of
relic gravitons and the smooth transition to the radiation
dominated era) but having a positive bulk viscous stress in order
to have an increasing matter entropy during all evolution of the cosmic time.

From Eqs.~(\ref{tt}) and~(\ref{rr}) we obtain
that the bulk viscous coefficient for this kind of fluid may be written as
\begin{eqnarray}\label{15}
\kappa \xi=\frac{2\dot{H}}{3H}+ (\gamma+1)H.
\end{eqnarray}
Notice that Eq.~(\ref{15}) implies that the
parametrization~(\ref{sup}) has a state parameter $\gamma=-1$. So
we can consider viscous cosmological models for which the state
parameter $\gamma \neq -1$, in other words we shall find a self--consistent solution for
the full set of Einstein field equations. The second term of~(\ref{15}) may be
positive and then we can have a non--negative bulk viscosity for
all cosmic evolution. We can see that this is possible for
expanding universes (for which $H>0$) with state parameter $\gamma
>-1$.

Now from the field equations~(\ref{tt}) and~(\ref{rr})
we have that
\begin{eqnarray}\label{Peffr}
\kappa P_{eff}= -3 H^2 - 2 \dot{H},
\end{eqnarray}
and then this new class
of viscous models will have the same effective
pressure~(\ref{PeffG}), which behaves as $P_{eff} \rightarrow
\rho/3$ for $t \rightarrow +\infty $.

The new class of solutions can be written as
\begin{eqnarray}\label{CS}
a(t)= \left( t+\sqrt{t^2+t_1^2}\right)^{1/2},\,\,\,\,\, \kappa
\rho(t)=\frac{3}{4 \left( t^2+t_1^2 \right)},
\end{eqnarray}
\begin{eqnarray}\label{CS1}
p=\gamma \rho, \,\,\,\,\, \kappa \xi(t)=
\frac{3(\gamma+1)\sqrt{t^2+t_1^2}-4t}{6 \left(t^2+t_1^2\right)}.
\end{eqnarray}
From here we get that the viscous pressure is given by
\begin{eqnarray}
\kappa \Pi=-3 \kappa H
\xi=\frac{4t-3(\gamma+1)\sqrt{t^2+t_1^2}}{4(t^2+t_1^2)^{3/2}},
\end{eqnarray}
and then $P_{eff}$ takes the form of Eq.~(\ref{PeffG}).

Now from Eq.~(\ref{CS1}) we can derive the general behavior of
$\xi$. Note that its numerator in general can be positive,
negative, or zero. It can be shown that, if
\begin{eqnarray}\label{root}
t_{root}=\frac{(\gamma+1)t_1}{\sqrt{\left(\frac{1}{3}-\gamma\right)\left(\gamma+\frac{7}{3}\right)}},
\end{eqnarray}
the bulk viscous stress is zero. This root is a real one if $-7/3
\leq\gamma \leq 1/3$. This means that, in this range of the state
parameter, the bulk viscous stress has a phase where $\xi$ is
positive (the matter entropy increases with time) and another
phase where it is negative (the matter entropy decreases with
time). In this case, if $-7/3 < \gamma < -1$, the value of
$t_{root}$ is negative and, if $-1 < \gamma < 1/3$, the value of
$t_{root}$ is positive. This is shown in Fig.\ref{fig159}. Note
that the solution studied in Ref.~\cite{Giovannini,Giovannini1}
lies in this range so, for $-7/3 \leq\gamma \leq 1/3$, we have a
class of Giovannini--like solutions with $\gamma \neq -1$.
\begin{figure}
\includegraphics{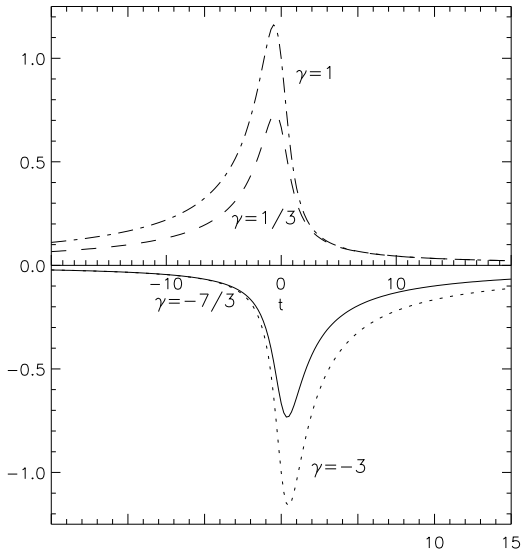}
\caption{\label{fig159a} Some curves for $\kappa\xi(t)$ given in
Eq.~(\ref{CS1}) are plotted. We set $t_1=1$ and
$\gamma=-3,-7/3,1/3,1$ as illustrative values.}
\end{figure}
The solutions for which the matter entropy always increases with
cosmic time lie out of the range $-7/3 \leq\gamma \leq 1/3$. In
this case the root~(\ref{root}) does not exist and then the bulk
viscosity is always negative or always positive. Effectively one
can show that, for $\gamma \leq -7/3$, the bulk viscous stress is
always negative, and for $\gamma \geq 1/3$ the bulk viscous stress
is always positive (see Fig.~\ref{fig159a}). 

In conclusion, any viscous cosmology with a constant barotropic
state parameter $\gamma \geq 1/3$ will have a positive bulk
viscous stress, and thus an increasing matter entropy during all
cosmic evolution. This new class of solutions has the following
asymptotic behavior at $t \rightarrow +\infty$: $\kappa \rho
\rightarrow 3/4t^2$, $\kappa P_{eff} \rightarrow 1/4t^2$, $\kappa
\xi \rightarrow (3 \gamma-1)/6t$ and always $p=\gamma \rho$. From
here we conclude that the radiative viscous fluid solution is more
physically acceptable than other ones since, for it, at first
order, we have $P_{eff} \rightarrow p$, $\xi \approx 0$ and always
$p=\rho/3$. So, it should be emphasized that if $t \to +\infty$
the scale factor expands as $a \to t^{1/2}$ while the equation of state 
of the local equilibrium pressure $p$ may be different from $\rho/3$. 
However, what counts is the effective pressure for which we have that
$P_ {eff} \to \rho/3$ as it should.  

Notice also that the sign of the
effective enthalpy, i.e. $\rho+P_{eff}$, plays an important role. Effectively,
we can see that the DEC is associated with Eq.~(\ref{DECViol}), which implies that for $\dot{H}>0$
DEC is violated, and for $\dot{H}<0$ DEC is preserved. On the other hand, as we have
stated above, the slopes of the energy spectra of relic gravitons are crucially related
to the sign of the effective enthalpy. The energy density of the created gravitons
increases with frequency if $\rho+P_{eff} \leq 0$, i.e. if the DEC is violated  ($\dot{H}>0$).  
So the considered here viscous cosmological models with a constant barotropic
state parameter $\gamma \geq 1/3$, violate the DEC only for a finite amount of time. In this early stage
($t<0$) the relic gravitons are created, and when the DEC is restored ($t>0$) the universe
exits to the standard radiation dominated stage. During all evolution of the cosmic time the bulk viscosity is positive.
In this way we have shown that the possible bulk viscous influence
on the slopes of energy spectra of relic gravitons makes sense
even without violation of the standard second law of
thermodynamics.

Lastly, notice that we have considered the same form for the
background metric~(\ref{GS}); thus all calculations reported in
Ref.~\cite{Giovannini} are preserved since they rely mostly on the
specific time dependence of the scale factor, rather than on the
bulk viscosity coefficient. Thus we come to the same conclusions
concerning the amplification induced by the background~(\ref{GS})
in the proper amplitude of the gravitational waves reported in
Ref.~\cite{Giovannini}. It is important to stress here that
these results imply that the existence of tilted spectra of relic gravitons
can be connected, in the framework of general relativity, with the 
violation of the DEC induced by a bulk viscosity compatible with the standard
second law of thermodynamics. Finally, going back to
the original Giovannini motivation~\cite{Giovannini}, we conclude that these
results indicate the possibility of the existence of growing spectra of gravitational
waves at frequencies of the interferometric devices which can put bounds on the possible 
violation of the DEC occurring in the early universe. It is interesting to note that growing
energy spectra can also be obtained in the framework of non-Einsteinian theories~\cite{Giovannini000,Giovannini2}.

\section{Acknowledgements}
The authors thank Paul Minning for carefully reading this
manuscript. This work was supported by CONICYT through Grant
FONDECYT N$^0$ 1051086 (MC, PM) and by Direcci\'on de Investigaci\'on de la
Universidad del B\'\i o--B\'\i o (MC). The financial support of
Escuela de Graduados of the Universidad de Concepci\'on is
acknowledged (PM).

\end{document}